\documentclass[12pt,preprint]{aastex}

\shorttitle{Current Helicities in Coronal Holes} \shortauthors{Yang,
Zhang, Li, \& Ding}

\begin{document}

\title{Vector Magnetic Fields and Current Helicities in Coronal Holes and Quiet Regions}

\author{Shuhong Yang\altaffilmark{}, Jun Zhang\altaffilmark{}, Ting Li\altaffilmark{}}

\affil{Key Laboratory of Solar Activity, National Astronomical
Observatories, Chinese Academy of Sciences, Beijing 100012, China}
\email{[shuhongyang;zjun;liting]@nao.cas.cn}

\author{Mingde Ding\altaffilmark{}}

\affil{Department of Astronomy, Nanjing University, Nanjing 210093,
China} \affil{Key Laboratory for Modern Astronomy and Astrophysics
(Nanjing University), Ministry of Education, Nanjing 210093, China}
\email{dmd@nju.edu.cn}

%\altaffiltext{}{}

\begin{abstract}

In the solar photosphere, many properties of coronal holes (CHs) are
not known, especially vector magnetic fields. Using observations
from \emph{Hinode}, we investigate vector magnetic fields, current
densities and current helicities in two CHs and compare them with
two normal quiet regions (QRs) for the first time. We find that, in
the CHs and QRs, the areas where large current helicities are
located are mainly co-spatial with strong vertical and horizontal
field elements both in shape and location. In the CHs, horizontal
magnetic fields, inclination angles, current densities and current
helicities are larger than those in the QRs. The mean vertical
current density and current helicity, averaged over all the observed
areas including the CHs and QRs, are approximately 0.008 A m$^{-2}$
and 0.005 G$^{2}$ m$^{-1}$, respectively. The mean current density
in magnetic flux concentrations where the vertical fields are
stronger than 100 G is as large as 0.012 $\pm$ 0.001 A m$^{-2}$,
consistent with that in the flare productive active regions. Our
results imply that the magnetic fields, especially the strong
fields, both in the CHs and QRs are nonpotential.

\end{abstract}


\section{Introduction}

Coronal holes (CHs) are low density and temperature regions in the
solar atmosphere (Munro \& Withbroe 1972). They appear as dark areas
observed with X-ray or EUV lines. CHs are always classified into
three categories according to their locations and lifetimes: polar,
nonpolar (isolated), and transient (Harvey \& Recely 2002). In
previous studies, many properties of CHs and their relationship with
magnetic fields have been investigated by many authors, such as
temperature variation (Wilhelm 2006; Zhang et al. 2007), element
abundance (Laming \& Feldman 2003), magnetic field evolution (Yang
et al. 2009a, b), and magnetic field structures (Meunier 2005; Zhang
et al. 2006; Tian et al. 2008).

In CHs, magnetic fields are predominated by one polarity and open
magnetic lines are concentrated (Bohlin 1977). Plasma escapes along
the open magnetic flux, giving rise to fast solar wind (Krieger et
al. 1973; Tu et al. 2005). CHs differ from the normal quiet Sun
regions mainly through the difference of magnetic structures, i.e.,
the field lines are mainly closed in the quiet Sun and open above
CHs (Altschuler et al. 1972). Hot gas is able to escape along open
field lines but is trapped in closed loops. The normal quiet Sun
appears brighter due to the radiation of the trapped gas. However,
the magnetic fields are not exclusively unipolar in CHs and
consequently CHs should contain also locally closed coronal loops
besides the open flux (Levine 1977). Wiegelmann \& Solanki (2004)
computed some properties of coronal loops. They found that high and
long closed loops are extremely rare, whereas short and low-lying
loops are almost as abundant in CHs as in the quiet Sun. This result
suggests an explanation for the relatively strong chromospheric and
transition region emission (many low-lying, short loops), but the
weak coronal emission (few high and long loops) in CHs.

Vector magnetic fields are very significant since they can provide
us plentiful information, such as electric current and current
helicity. In a common consensus, the free energy is stored in the
stressed, nonpotential magnetic fields. Electric current and current
helicity are two parameters which are always used to characterize
magnetic nonpotentiality (Moreton \& Severny 1968; Abramenko et al.
1996; Wang 1996; Zhao et al. 2009; Su et al. 2009). Electric current
density \textbf{\emph{J}} is calculated as
\begin{equation}
\textbf{\emph{J}} =\frac{1}{\mu_{0}}\emph{\textbf{$\nabla\times$B}},
\end{equation}
where $\mu_{0}$ is the magnetic permeability in vacuum
(4$\pi$$\times$10$^{-3}$ G m A$^{-1}$). By definition, current
helicity density \emph{h}$_{c}$ is derived as
\begin{equation}
\emph{h}_{c}=\emph{\textbf{B$\cdot$$\nabla\times$B}}.
\end{equation}
The larger the deviation from the potential fields is, the more the
energies that can power solar activity are. Therefore, electric
current and current helicity also serve as measures of the
productivity of solar activity.

Electric current and current helicity have been intensively studied
in active regions (ARs), while quantitative study is quite rare in
CHs. This is mainly due to the relative weak fields in CHs and the
former instruments are not capable of achieving reliable vector
field observations. Fortunately, \emph{Hinode} (Kosugi et al. 2007)
provides us wonderful space-based measurements which are vastly
superior to previous data in resolution and sensitivity. The direct
purpose of this study is to compare vector magnetic fields, current
densities and current helicities in CHs and quiet regions (QRs) with
\emph{Hinode} observations. We describe the observations and data
reduction in Section 2. The results are presented in Section 3, and
the conclusions and discussion in Section 4.

%\clearpage

\section{Observations and data reduction}

The data used here were carried out with the Spectro-Polarimeter
(SP; Lites et al. 2001) in the Solar Optical Telescope (SOT;
Ichimoto et al. 2008; Shimizu et al. 2008; Suematsu et al. 2008;
Tsuneta et al. 2008) instrument aboard \emph{Hinode}. The SP
provides Stokes \emph{I}, \emph{Q}, \emph{U}, and \emph{V} profiles
of two Fe lines at 630.15 nm and 630.25 nm with wavelength sampling
of 21.6 m{\AA} in four modes (normal, fast, dynamics and deep maps).
We adopt the observations of two CHs (named CH1 and CH2 here) and
two non-coronal hole QRs (QR1 and QR2) taken in the fast map mode.
For each raster slit, the integrated exposure time is 3.2 s and the
pixel sampling along the slit is 0{\arcsec}.32. The scan is in the
east-west direction with a scanning step of 0{\arcsec}.30. CH1 was
observed from 11:42:25 UT to 12:45:36 UT on November 30, 2008, and
QR1 from 17:34:06 UT to 18:37:18 UT on April 12, 2007. Both of them
have a field-of-view (FOV) of 300{\arcsec}${\times}$162{\arcsec}.
They are located almost at the disk center. CH2 was taken from
01:36:29 UT to 02:08:49 UT on November 12, 2007, and QR2 from
11:35:09 UT to 12:21:47 UT on April 13, 2007. Both of them have a
FOV of 151{\arcsec}${\times}$162{\arcsec}. They are centered at
about 10$\degr$ from the disk center.

The SP data are calibrated and inverted at the Community
Spectro-polarimetric Analysis Center (CSAC;
\url{http://www.csac.hao.ucar.edu/}). Vector magnetic fields are
retrieved using the inversion techniques based on the assumption of
the Milne-Eddington atmosphere model (Kubo et al. 2007; Yokoyama, T.
2010, in preparation). In the inversion, a non-linear least-squares
fitting technique is used to fit analytical Stokes profiles to the
observed profiles. Values of 36 parameters are returned from the
inversion, including the three components of magnetic field (field
strength $B$, field inclination $\gamma $, field azimuth $\phi$),
the stray light fraction $\alpha$, and so on. In the vector field
measurements based on the Zeeman Effect, there exists a 180$\degr$
ambiguity in determining the field azimuth. Various algorithms have
been developed to resolve the ambiguity, but it is also difficult to
make a complete removal (Metcalf et al. 2006; Semel \& Skumanich
1998). As pointed out by Metcalf et al. (2006), ``the methods which
minimize some measure of the vertical current density in conjunction
with minimizing an approximation for the fields' divergence show the
most promise." In this study, we use the improved Nonpotential
Magnetic Field Calculation (NPFC) method developed by Georgoulis
(2005) to disambiguate the azimuth angles.

The vector magnetic field \emph{\textbf{B}} is shown by longitudinal
field strength \emph{B}\emph{cos}$\gamma$ and transverse field
strength \emph{B}\emph{sin}$\gamma$ without involving filling
factor, where \emph{B} is the intrinsic strength of the magnetic
field and $\gamma$ the inclination angle with respect to the
line-of-sight direction. The transverse field can be decomposed into
$B_{\xi}$=\emph{B}\emph{sin}$\gamma$\emph{cos}$\phi$ and
$B_{\eta}$=\emph{B}\emph{sin}$\gamma$\emph{sin}$\phi$ in the image
plane, where $\phi$ is the field azimuth angle. Then we transform
the vector magnetic field at each pixel to heliographic components
($B_{x}, B_{y}, B_{z}$) according to the formulae given by Gary \&
Hagyard (1990). Besides, geometric mapping of the magnetograms in
the image plane into the heliographic plane is also performed.
Vector magnetic fields in the photosphere allow us to compute the
vertical component $J$$_{z}$ of electric current \emph{\textbf{J}}
in the photospheric plane where \emph{z} = 0. $J$$_{z}$, i.e.,
\begin{equation}
J_{z}(x, y, z=0) =\frac{1}{\mu_{0}} (\frac{ {\partial}B_{y}
}{{\partial}x} - \frac{ {\partial}B_{x} }{{\partial}y})|_{z=0},
\end{equation}
is calculated by finite differences. Then we obtain the current
helicity density $h_{c}$ as,
\begin{equation}
h_{c}(x, y, z=0) ={\mu_{0}}B_{z}J_{z}|_{z=0}.
\end{equation}
Here, $h_{c}$ is just part of the current helicity density since
only the vertical current density can be calculated.

We determine the noise level of vertical fields by calculating
1$\sigma$ standard deviation of weak fields within intranetwork
regions and find it is about 4 G. The noise level of horizontal
fields determined in the similar way is about 35 G. For current
density, the noise level (0.005 A m$^{-2}$) is estimated from the
standard deviation of them in the pixels where the horizontal fields
are weaker than 35 G, similar to the method of de La
Beaujardi\`{e}re et al. (1993), Wang et al. (1996) and Wheatland
(2000). For current helicity density, the noise level (0.004 G$^{2}$
m$^{-1}$) is calculated from the standard deviation of them over the
pixels where either the vertical or the horizontal fields are weaker
than their noise levels.

%\clearpage

\section{Results}

CH1 and QR1 are located at the disk center, while CH2 and QR2 have
nearly the same heliocentric angle. We investigate the vector
magnetic fields, current densities and current helicities of each
pair of targets, respectively, and the unsigned quantities are used
throughout.

%\clearpage

\subsection{Comparison between CH1 and QR1}

CH1 is dominated by the positive polarity as shown in the upper
panel in Figure 1. The dash-dotted curve encloses CH1 area the
pixels within which are considered only. We plot the probability
density functions (PDFs) of the vertical fields in CH1 and QR1 in
Figure 2 (upper panel). Not unexpectedly, CH1 fields have a
generally imbalanced PDF between the positive and negative
polarities. The positive fields, which are the predominant polarity,
have a higher PDF than the negative. We define the magnetic flux
imbalance parameter $\rho$ as
\begin{equation}
\rho =\frac{ \Sigma \Phi (i,j) }{ \Sigma | \Phi (i,j) | },
\end{equation}
where $\Phi$$(i,j)$ is the vertical magnetic flux at pixel $(i,j)$.
As presented in the lower panel, the flux imbalance in CH1 is about
0.65, much higher than that in QR1 (0.25). Figures 3 and 4 display
the appearance of horizontal magnetic fields and the distribution of
derived vertical current helicities in CH1 and QR1, respectively.
When we only consider the areas where unsigned vertical magnetic
fields are stronger than 4 G and horizontal fields are stronger than
35 G, the mean vertical magnetic field, horizontal field, current
density and current helicity in CH1 are 33.27 $\pm$ 0.01 G, 96.64
$\pm$ 0.07 G, 0.00888 $\pm$ 0.00001 A m$^{-2}$, and 0.00492 $\pm$
0.00001 G$^{2}$ m$^{-1}$, while those in QR1 are 39.44 $\pm$ 0.01 G,
89.20 $\pm$ 0.06 G, 0.00839 $\pm$ 0.00001 A m$^{-2}$, and 0.00558
$\pm$ 0.00001 G$^{2}$ m$^{-1}$, respectively. Here, the
uncertainties of the means are calculated according to Equation
(4.14) in Bevington \& Robinson (2003).

We examine the distribution of current helicities (Figure 4) in CH1
and QR1 and find that the large current helicity concentrations are
mainly co-spatial with the strong vertical and horizontal field
elements (Figures 1 and 3). The region highlighted with square ``1"
in Figure 4 is investigated in detail. At the center of this region,
large current helicities are distributed. In Figures 1$-$3, the
contours are vertical magnetic fields at $\pm$ 100 G levels. At the
places where large current helicities are located, the vertical and
horizontal magnetic fields (squares ``1" in Figures 1 and 3) are
much stronger than the surrounding areas. We take 100 G as the
general separation of strong and weak magnetic fields. The mean
vertical field strength within the contour is 584.1 $\pm$ 0.2 G and
the mean horizontal field strength is 162.3 $\pm$ 1.3 G. The
corresponding average unsigned current density and current helicity
are 0.0130 $\pm$ 0.0002 A m$^{-2}$ and 0.0962 $\pm$ 0.0002 G$^{2}$
m$^{-1}$, respectively. In QR1, we select another sub-region (square
``2" in Figure 4) where large current helicities are concentrated.
Similar to region ``1" in CH1, the areas with large current
helicities also correspond with strong vertical and horizontal
fields. The mean vertical field, horizontal field, electric current
and current helicity within the contour curve in region ``2" are
593.8 $\pm$ 0.1 G, 176.7 $\pm$ 1.3 G, 0.0131 $\pm$ 0.0002 A
m$^{-2}$, and 0.0988 $\pm$ 0.0001 G$^{2}$ m$^{-1}$.

To examine whether these magnetic properties change with vertical
magnetic field strength, we calculate the mean values of horizontal
magnetic fields, field inclination angles, current densities, and
current helicities above some thresholds of unsigned vertical
magnetic fields in CH1 and QR1 (Figure 5). With the increase of
threshold from weak vertical fields to about 100 G, the mean value
of horizontal fields in CH1 increases sharply from 97 G to 145 G, as
shown in Figure 5a. When the vertical fields exceed 100 G, the
horizontal fields exhibit a slight increase to the level of 180 G.
The relationship between vertical fields and horizontal fields in
QR1 is similar to that in CH1 except for about 15 G weakness. The
mean inclination angle of the magnetic lines in CH1 is about
75$\degr$ for all the fields (Figure 5b). Then it decreases quickly
to about 38$\degr$ with the increase of vertical fields in the range
from weak to about 100 G. When the vertical fields are stronger than
100 G, the inclination angles gradually decrease to less than
8$\degr$. While the mean angles in QR1 are about 5$\degr$ smaller
that those in CH1. In CH1, the current densities increase with the
vertical fields and then stay at the level of 0.014 A m$^{-2}$, as
presented in Figure 5c. While the average current density in QR1 is
about 0.012 A m$^{-2}$. The difference between them is relatively
significant considering their uncertainties. The current helicities
in CH1 reveal a general trend of increase when the vertical fields
increase from weak fields to kilogauss (Figure 5d). We notice that
the current helicities in QR1 exhibit similar variation trends to
those in CH1 with smaller means.

%\clearpage

\subsection{Comparison between CH2 and QR2}

CH2 is an enhanced network field region with predominance of
negative polarity (top left panel in Figure 6) and QR2 a normal
quiet Sun with mixed-polarity (top right). We also examine the PDFs
of their vertical magnetic fields. Similar to those in CH1 and QR1,
the PDF in CH2 is clearly asymmetric while the PDF in QR2 shows
balanced positive and negative magnetic fields (upper panel in
Figure 7). Compared with CH1 in Figure 2, CH2 shows a much more
serious imbalance between the opposite polarities (lower panel in
Figure 7). The flux imbalance $\rho$ in CH2 is as high as $-$0.8 and
that in QR2 is only about $-$0.1. The mean vertical magnetic field,
horizontal field, current density and current helicity in CH2 are
41.21 $\pm$ 0.01 G, 86.94 $\pm$ 0.07 G, 0.00808 $\pm$ 0.00001 A
m$^{-2}$, and 0.00543 $\pm$ 0.00001 G$^{2}$ m$^{-1}$, while those in
QR2 are 33.34 $\pm$ 0.01 G, 84.42 $\pm$ 0.08 G, 0.00796 $\pm$
0.00001 A m$^{-2}$, and 0.00429 $\pm$ 0.00001 G$^{2}$ m$^{-1}$,
respectively.

In Figure 6, the bottom panels are corresponding current helicity
distributions in CH2 and QR2 scaled between $\pm$ 0.03 A m$^{-2}$.
From their general appearance, we can see that large current
helicities are also mainly co-spatial with strong magnetic fields
both in shape and location. We select two areas (highlighted with
windows ``3" and ``4") containing large current helicity patches.
When we take the pixels with vertical fields stronger than 100 G
into consideration, the mean vertical field strength in region ``3"
is as strong as 733.1 $\pm$ 0.2 G, and the mean horizontal field,
current density and current helicity are 152.8 $\pm$ 1.6 G, 0.0118
$\pm$ 0.0002 A m$^{-2}$, and 0.1153 $\pm$ 0.0002 G$^{2}$ m$^{-1}$,
respectively. While these four parameters in region ``4" are 511.3
$\pm$ 0.3 G, 109.0 $\pm$ 2.2 G, 0.0121 $\pm$ 0.0003 A m$^{-2}$, and
0.0824 $\pm$ 0.0003 G$^{2}$ m$^{-1}$, respectively.

We also compare the variations of horizontal fields, inclination
angles, current densities and current helicities versus the
thresholds of unsigned vertical magnetic fields in CH2 and QR2. The
horizontal fields increase both in CH2 and QR2 before the vertical
field threshold reaches 100 G (Figure 8a). Then the horizontal
fields in CH2 increase slowly from 120 G to 150 G, while they mainly
maintain at the 115 G level in QR2. The inclination angles in both
CH2 and QR2 decrease with the increase of vertical fields in a
similar trend (Figure 8b). The inclination angles finally approach
to 5$\degr$. The trend of current density variation in CH2 is
consistent with that in QR2 (Figure 8c). They are similar to those
in CH1 and QR1 shown in Figure 5c. When the threshold of vertical
fields is several hundred Gauss, the mean of current densities in
CH2 is larger than that in QR2. The current helicities in both CH2
and QR2 reveal a general trend of increase when the vertical fields
increase from weak fields to kilo-Gauss though the rise slows down
(Figure 8d). Generally, the current densities in CH2 are larger than
those in QR2.

\subsection{Comparison between the CHs and QRs}

Figure 5 indicates the means of the horizontal magnetic fields,
inclination angles, current densities, and current helicities in CH1
are larger than those in QR1. While Figure 8 exhibits that in CH2,
only the horizontal magnetic fields and current helicities are
stronger than QR2, and the inclination angles and current densities
in them are the same for the stronger fields. To enlarge sample, we
combine two CHs (CH1 and CH2) and two QRs (QR1 and QR2),
respectively. We compare the parameters between the CHs and QRs
(Figure 9), and find that horizontal magnetic fields, inclination
angles, current densities and current helicities in the CHs are
larger than those in the QRs.

%\clearpage

\section{Conclusions and discussion}

Using the \emph{Hinode}$/$SP data, we investigate vector magnetic
fields, current densities and current helicities in two CHs (CH1 and
CH2), and compare them with two QRs (QR1 and QR2). To our knowledge,
this comparison has not been done using vector field measurements
before. We find that in the areas both in the CHs and QRs where
large current helicities are concentrated, there are strong vertical
and horizontal field elements and they are mainly co-spatial with
each other in shape and location. In the CHs, horizontal magnetic
fields, inclination angles, current densities and current helicities
are larger than those in the QRs. The mean values and their
uncertainties of five parameters in the areas where the unsigned
vertical magnetic fields are stronger than 4 G and horizontal fields
than 35 G are listed in Table 1. Averaged over the observed CHs and
QRs, the means of vertical magnetic fields, horizontal fields,
inclination angles, current densities and current helicities are
approximately 37 G, 90 G, 75$\degr$, 0.008 A m$^{-2}$ and 0.005
G$^{2}$ m$^{-1}$, respectively.

The mean vertical magnetic strength in CH2 is 41 G, stronger than
that in CH1 (33 G). The reason is that CH2 is located at an enhanced
network region. According to the MDI Synoptic Chart, the location of
CH2 is $\sim$250$\degr$ (Carrington longitude). At the same site,
there was a small mature AR on 17 October 2007, which has been
studied by Murray et al. (2010). We think that the dispersion of the
AR's negative polarity leads to the strong fields in CH2. The
average vertical magnetic field in QR2 is 33 G, somewhat weaker than
that in QR1 (39 G). It may be caused by the magnetic field
fluctuation of the quiet Sun in the scale of FOV of our data. As
revealed in our results, the CHs are dominated by one polarity,
while
magnetic fluxes in the QRs are generally balanced in sign. %However the
%differences between CHs and QRs in the aspects of vector magnetic
%fields, current densities and current helicities are insignificant
%since they are buried by error bars.

There is also another way to estimate whether the difference or
similarity is significant, i.e. using the uncertainties of the
vector field parameters from the inversion. But this method is not
available, because ``some of the errors of the inversion parameters
are wrong and the problem can not be fixed any time soon," as
pointed out by Lites (private communication). ``However, the values
for the inversion parameters themselves are OK."

Flaring ARs carry more current than simple ones which are closer to
potential. According to McIntosh (1990), the majority of regions are
simple. As presented in our results, both in the CHs and QRs, large
current helicity areas are almost located in the regions with strong
vertical and horizontal magnetic fields.  The mean current density
in magnetic concentrations where the vertical fields are stronger
than 100 G is as large as 0.012 $\pm$ 0.001 A m$^{-2}$, consistent
with that ($\sim$ 0.01 A m$^{-2}$) in the flare productive ARs (de
La Beaujardi\`{e}re et al. 1993; Leka et al. 1993; Wang et al. 1994;
Wang et al. 1996; Zhang 2001; Deng et al. 2001; Leka \& Barnes 2003;
Liu et al. 2007; Li et al. 2009). These results imply that the
photospheric magnetic fields, especially the strong fields, in the
CHs and QRs are nonpotential.

Since only two pairs of CHs and QRs are investigated here, more
observations will be taken into consideration to check the
similarities and differences between CHs and QRs.

%\clearpage

\acknowledgments {We are grateful to the anonymous referee for the
constructive comments on the manuscript. We thank Profs. J. X. Wang
and H. Q. Zhang and Dr. Y. Gao for their useful discussions.
\emph{Hinode} is a Japanese mission developed and launched by
ISAS/JAXA, with NAOJ as domestic partner and NASA and STFC (UK) as
international partners. It is operated by these agencies in
cooperation with ESA and NSC (Norway). \emph{Hinode} SOT/SP
Inversions were conducted at NCAR under the framework of the
Community Spectro-polarimetric Analysis Center (CSAC;
\url{http://www.csac.hao.ucar.edu/}). This work is supported by the
National Natural Science Foundations of China (40890161, 11025315,
41074123 and 11003024), the CAS Project KJCX2-YW-T04, the National
Basic Research Program of China under grant 2011CB811400.

 %This work is supported by the National Natural
%Science Foundations of China (G40890161, 10703007, and 10733020),
%the CAS Project KJCX2-YW-T04, the National Basic Research Program of
%China under grant G2006CB806303, and the Young Researcher Grant of
%National Astronomical Observatories, Chinese Academy of Sciences.

{}

\clearpage

\begin{deluxetable}{cccccc}
%\tabletypesize{\scriptsize}
\tablecaption{Mean values and their uncertainties of five parameters
in the areas with unsigned vertical magnetic fields stronger than 4
G and horizontal fields stronger than 35 G. \label{tab}}
\tablewidth{0pt} \tablehead{ \colhead{ } & \colhead{Vertical field}
& \colhead{Horizontal field} & \colhead{Inclination} &
\colhead{Current density} &
\colhead{Current helicity} \\
\colhead{ } & \colhead{(G)} & \colhead{(G)} & \colhead{(Degree)} &
\colhead{(A m$^{-2}$)} & \colhead{(G$^{2}$ m$^{-1}$)}} \startdata

CH1 & 33.27 $\pm$ 0.01 & 96.64 $\pm$ 0.07 & 77.14 $\pm$ 0.03  & 0.00888 $\pm$ 0.00001 & 0.00492 $\pm$ 0.00001 \\
QR1 & 39.44 $\pm$ 0.01 & 89.20 $\pm$ 0.06 & 74.98 $\pm$ 0.02  & 0.00839 $\pm$ 0.00001 & 0.00558 $\pm$ 0.00001 \\
CH2 & 41.21 $\pm$ 0.01 & 86.94 $\pm$ 0.07 & 73.41 $\pm$ 0.03  & 0.00808 $\pm$ 0.00001 & 0.00543 $\pm$ 0.00001 \\
QR2 & 33.34 $\pm$ 0.01 & 84.42 $\pm$ 0.08 & 74.93 $\pm$ 0.03  & 0.00796 $\pm$ 0.00001 & 0.00429 $\pm$ 0.00001 \\

\enddata
\end{deluxetable}
\clearpage

\begin{figure}
\centering
\includegraphics
[clip,angle=0,scale=0.85]{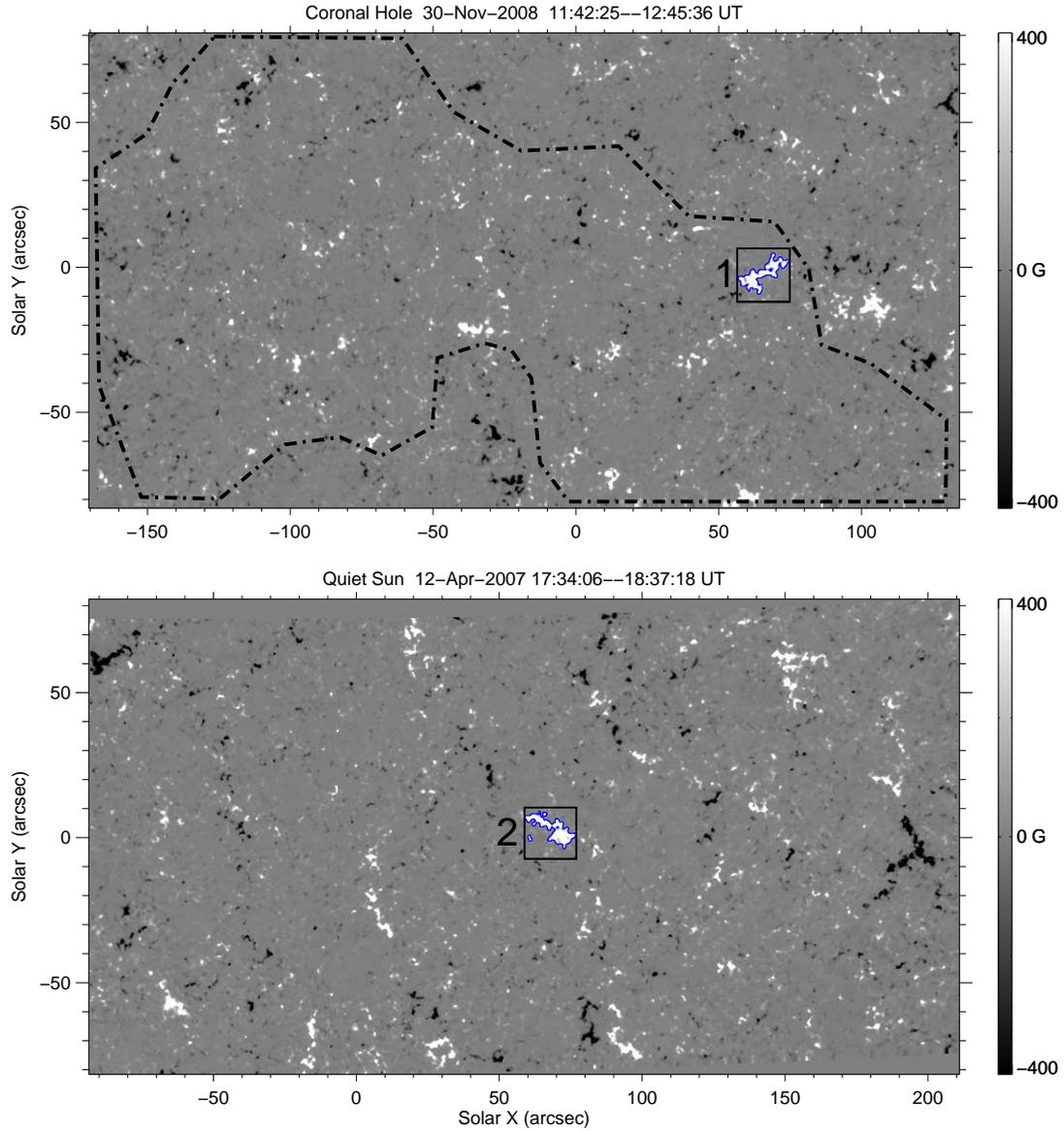} \caption{Overview of the vertical
magnetic fields in CH1 (upper panel) and QR1 (lower panel). The
dash-dotted curve delineates the CH boundary. Squares ``1" and ``2"
outline two sub-regions that are investigated in detail. The
contours in two squares are vertical magnetic fields at $\pm$ 100 G
levels. \label{fig}}
\end{figure}
\clearpage

\begin{figure}
\centering
\includegraphics
[bb=132 227 426 690,clip,angle=0,scale=1.]{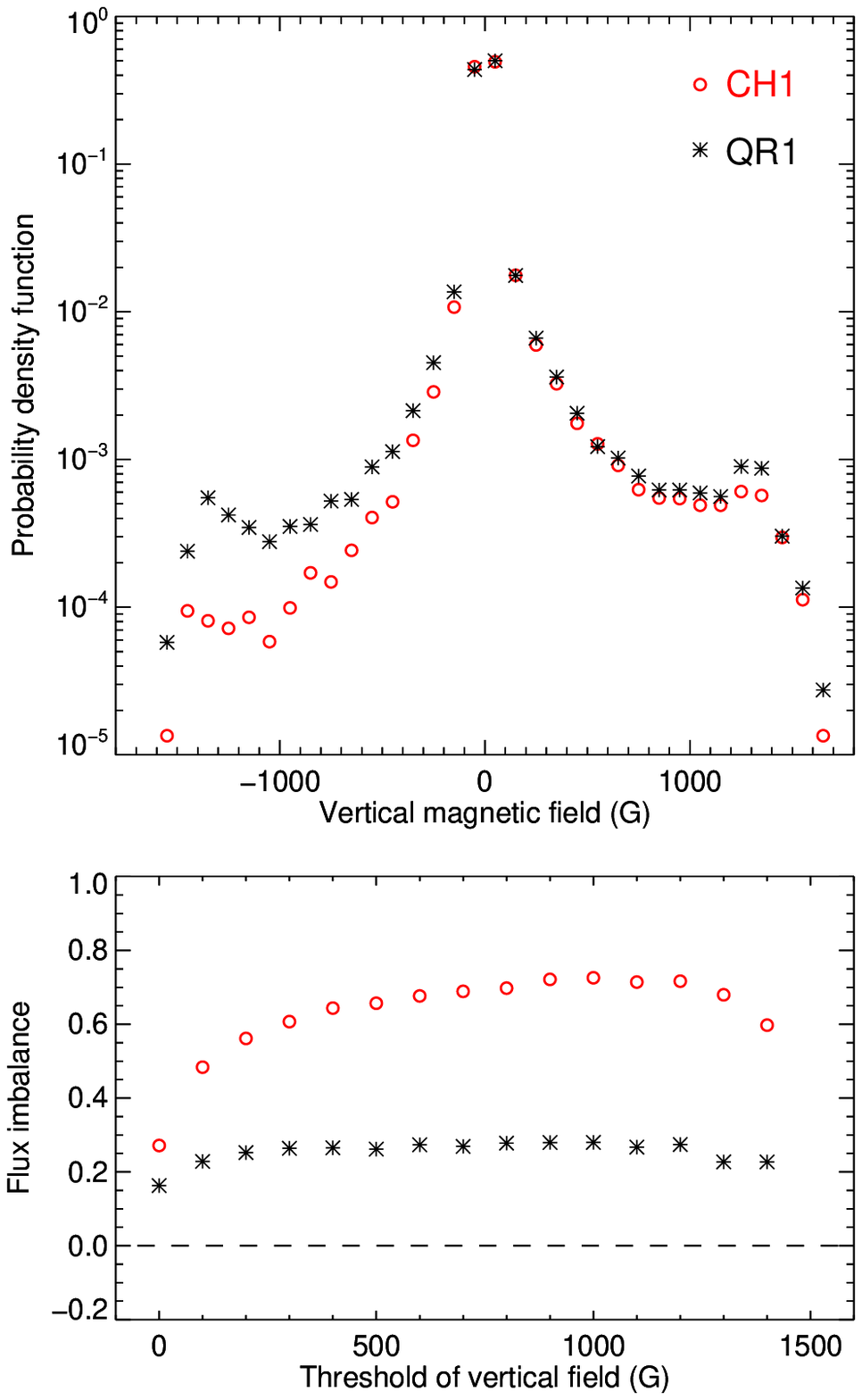}
\caption{Probability density function of the vertical magnetic
fields (upper panel) and flux imbalance (lower panel) in CH1 and
QR1. \label{fig}}
\end{figure}
\clearpage

\begin{figure}
\centering
\includegraphics
[clip,angle=0,scale=0.85]{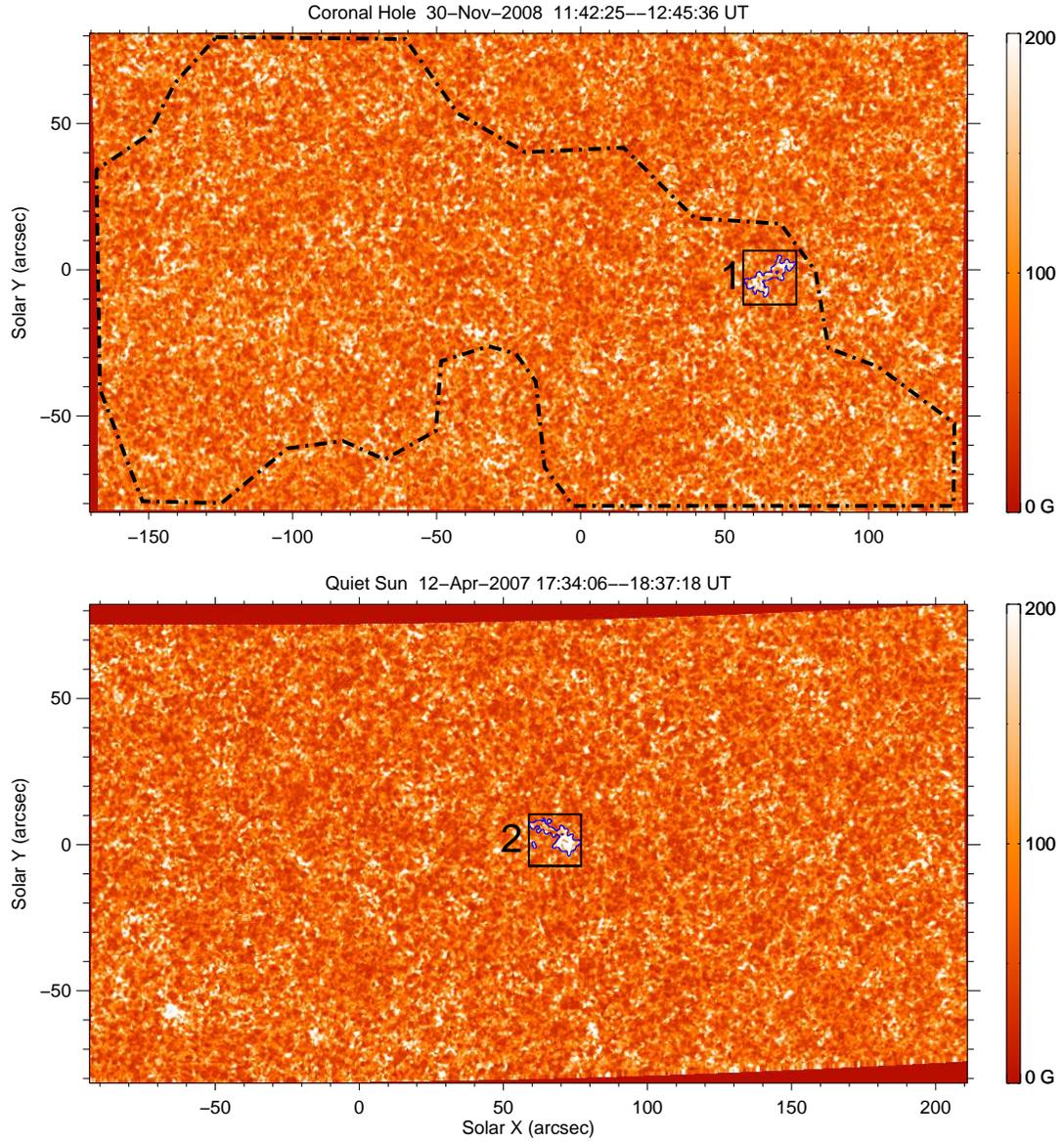} \caption{Appearance of the
horizontal magnetic fields corresponding to Figure 1. The
dash-dotted curve, squares, and contours have the same meanings as
those in Figure 1. \label{fig}}
\end{figure}
\clearpage

\begin{figure}
\centering
\includegraphics
[clip,angle=0,scale=0.85]{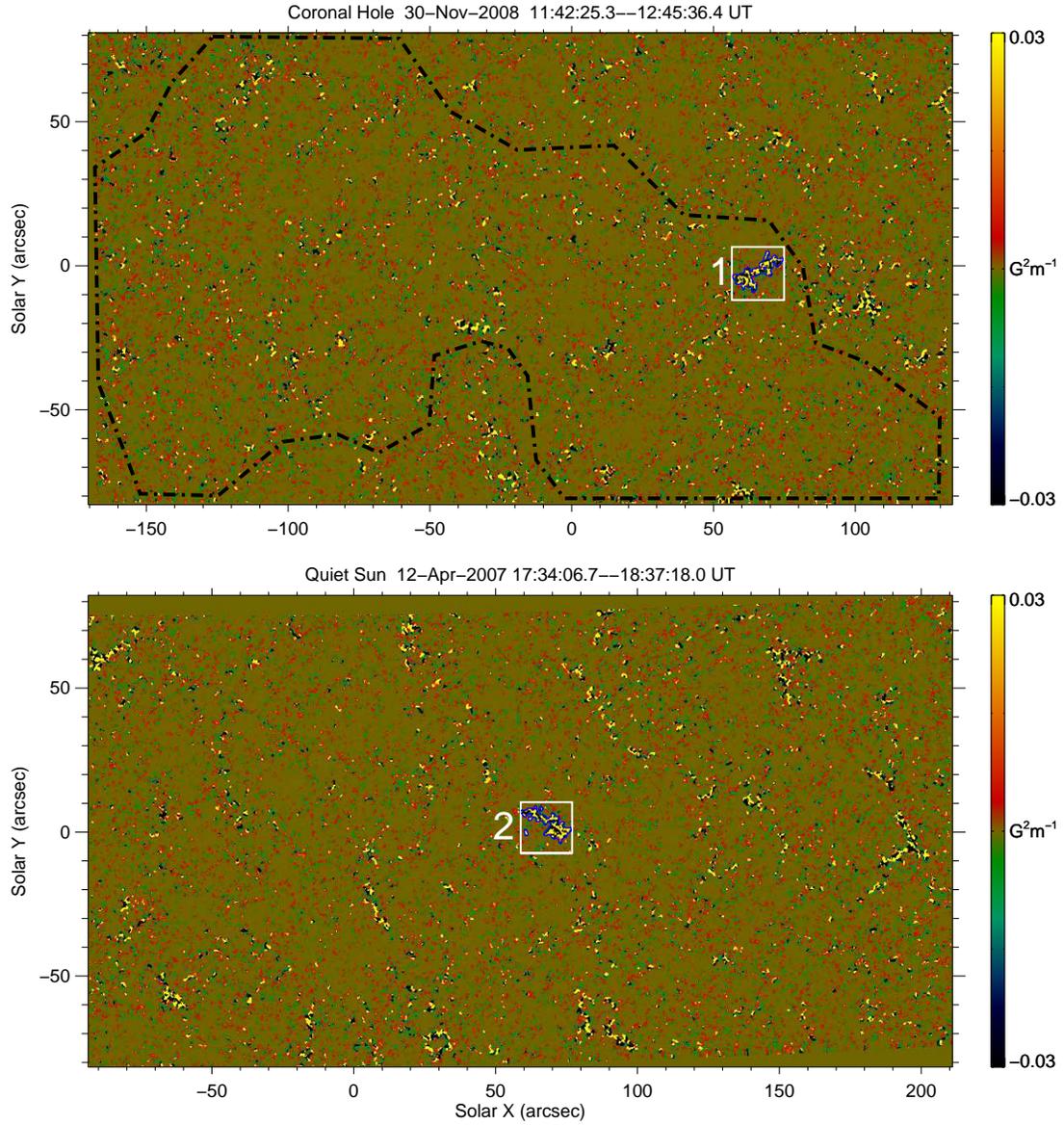} \caption{Distribution of the
vertical current helicities in CH1 and QR1. The dash-dotted curve,
squares, and contours have the same meanings as those in Figure 1.
\label{fig}}
\end{figure}
\clearpage

\begin{figure}
\centering
\includegraphics
[clip,angle=0,scale=0.95]{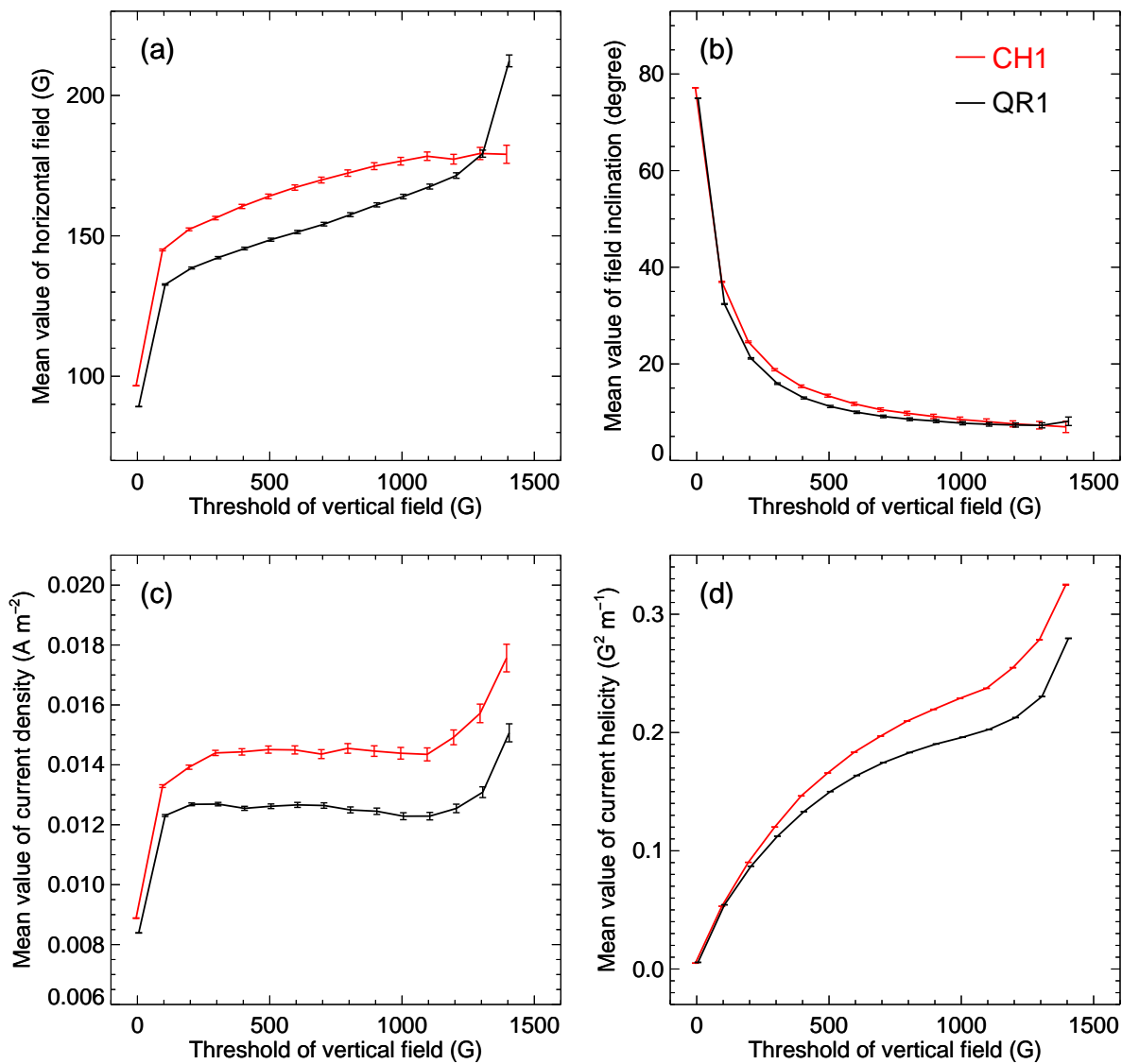} \caption{Variations of the
horizontal magnetic fields (a), field inclinations (b), current
densities (c), and current helicities (d) versus the thresholds of
unsigned vertical magnetic fields in CH1 and QR1. Each error bar
represents the standard deviation of the mean value. \label{fig}}
\end{figure}
\clearpage

\begin{figure}
\centering
\includegraphics
[bb=58 173 544 658,clip,angle=0,scale=0.85] {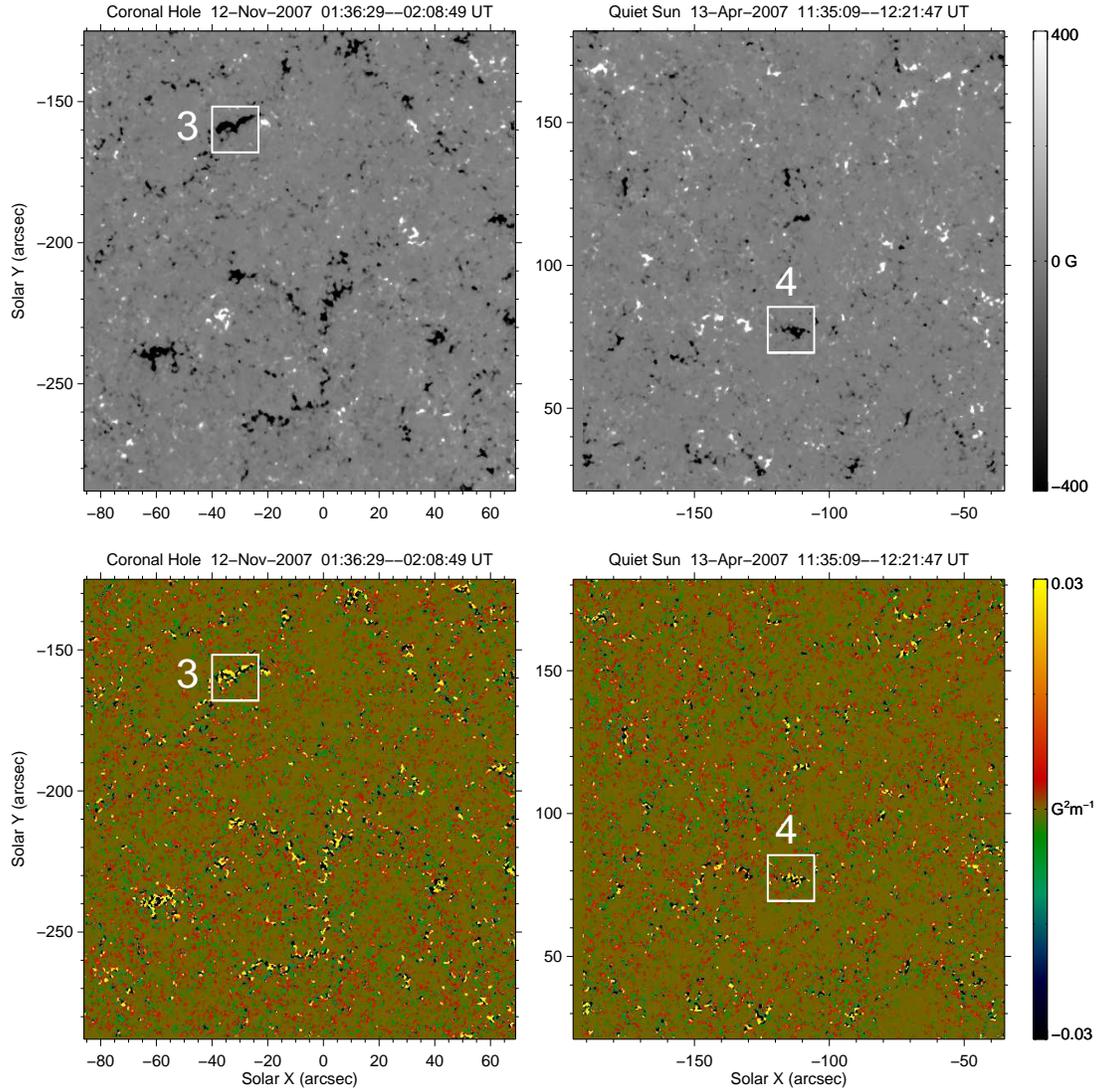}
\caption{Vertical magnetograms in CH2 (top left) and QR2 (top right)
and corresponding current helicity distributions (bottom panels).
Windows ``3" and ``4" highlight two areas that are studied in
detail. \label{fig}}
\end{figure}
\clearpage

\begin{figure}
\centering
\includegraphics
[bb=132 227 426 690,clip,angle=0,scale=0.95]{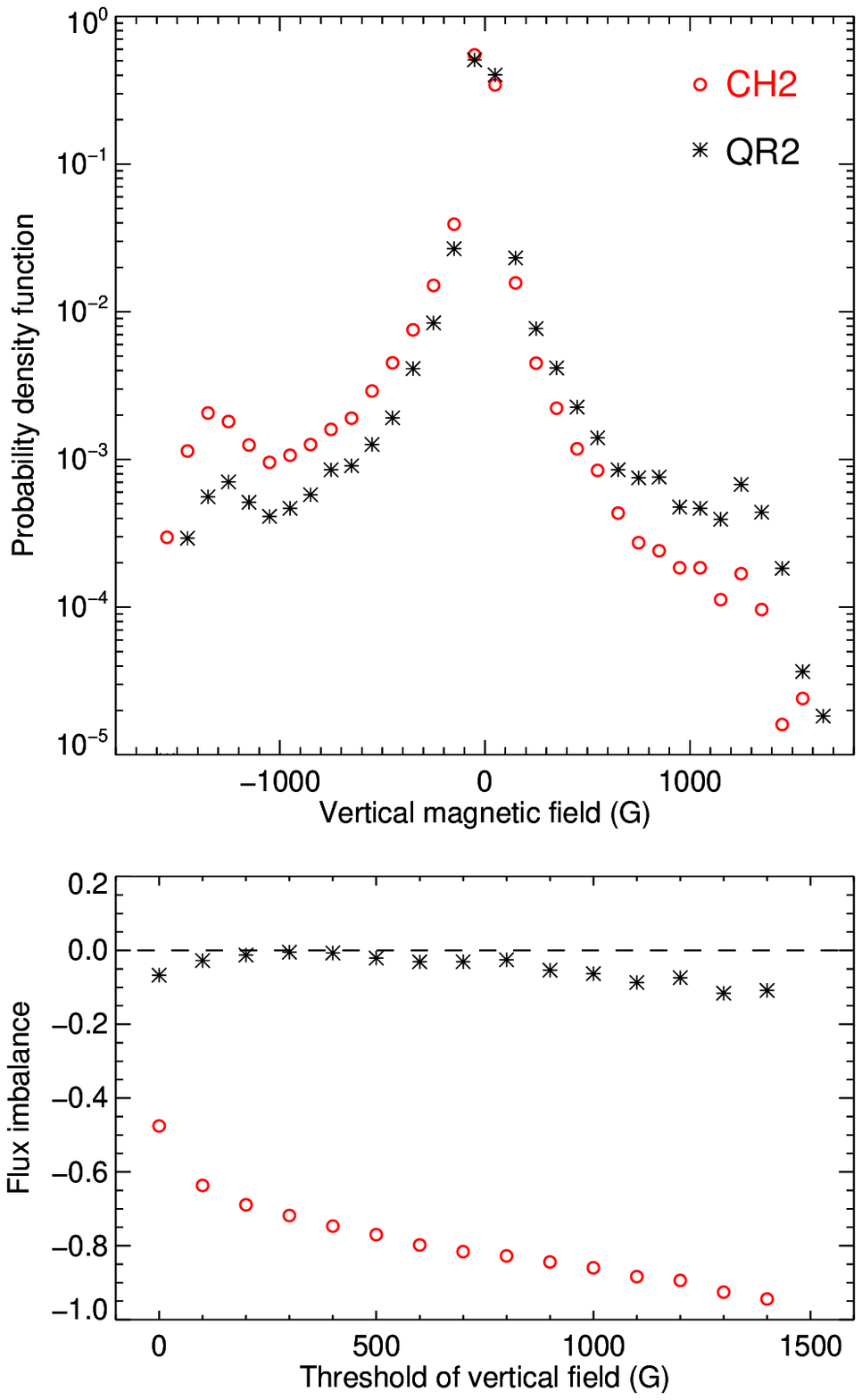}
\caption{Probability density function of the vertical magnetic
fields (upper panel) and flux imbalance (lower panel) in CH2 and
QR2. \label{fig}}
\end{figure}
\clearpage

\begin{figure}
\centering
\includegraphics
[clip,angle=0,scale=0.95]{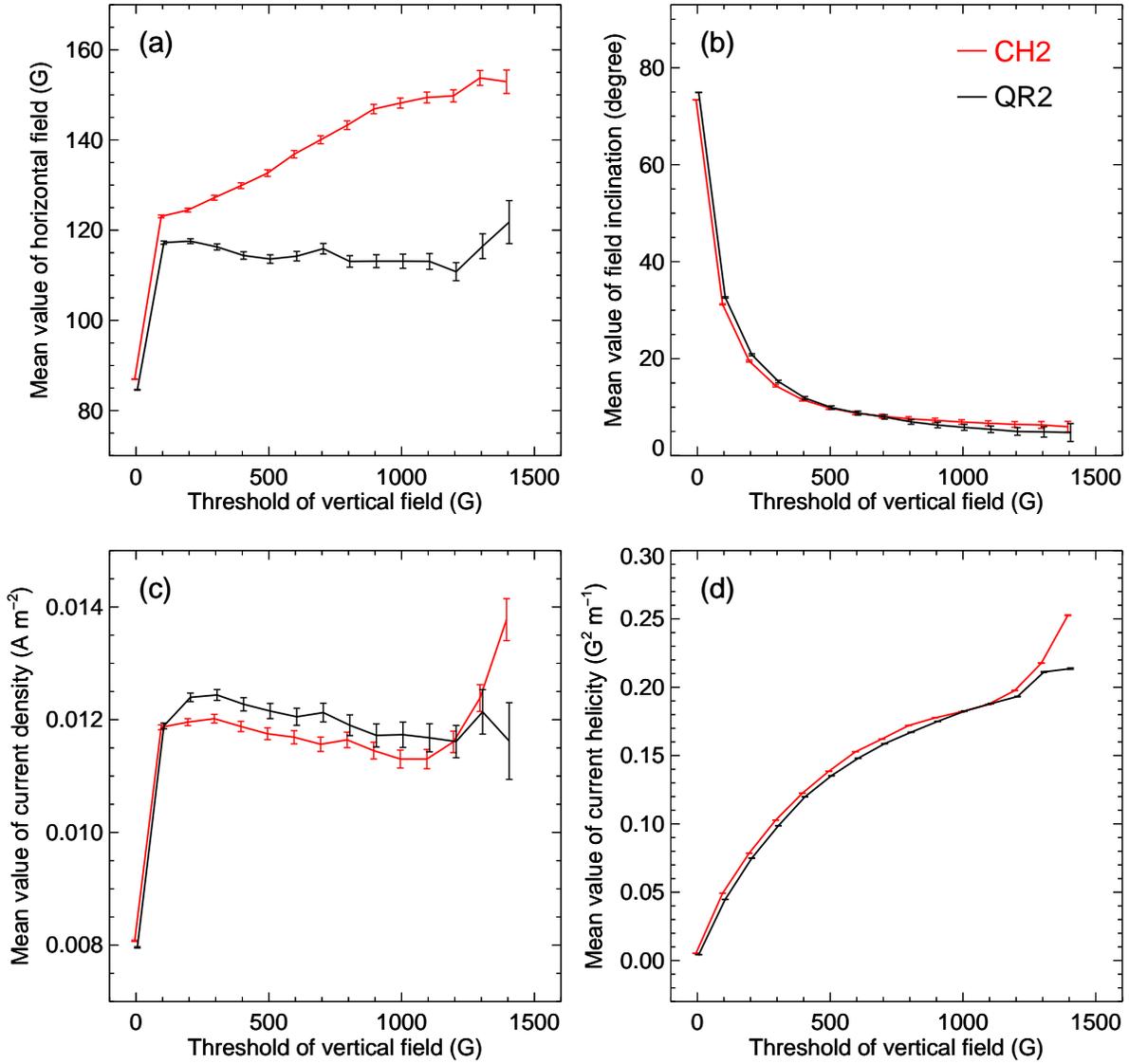} \caption{Similar to Figure 5 but
for CH2 and QR2. \label{fig}}
\end{figure}
\clearpage

\begin{figure}
\centering
\includegraphics
[clip,angle=0,scale=0.95]{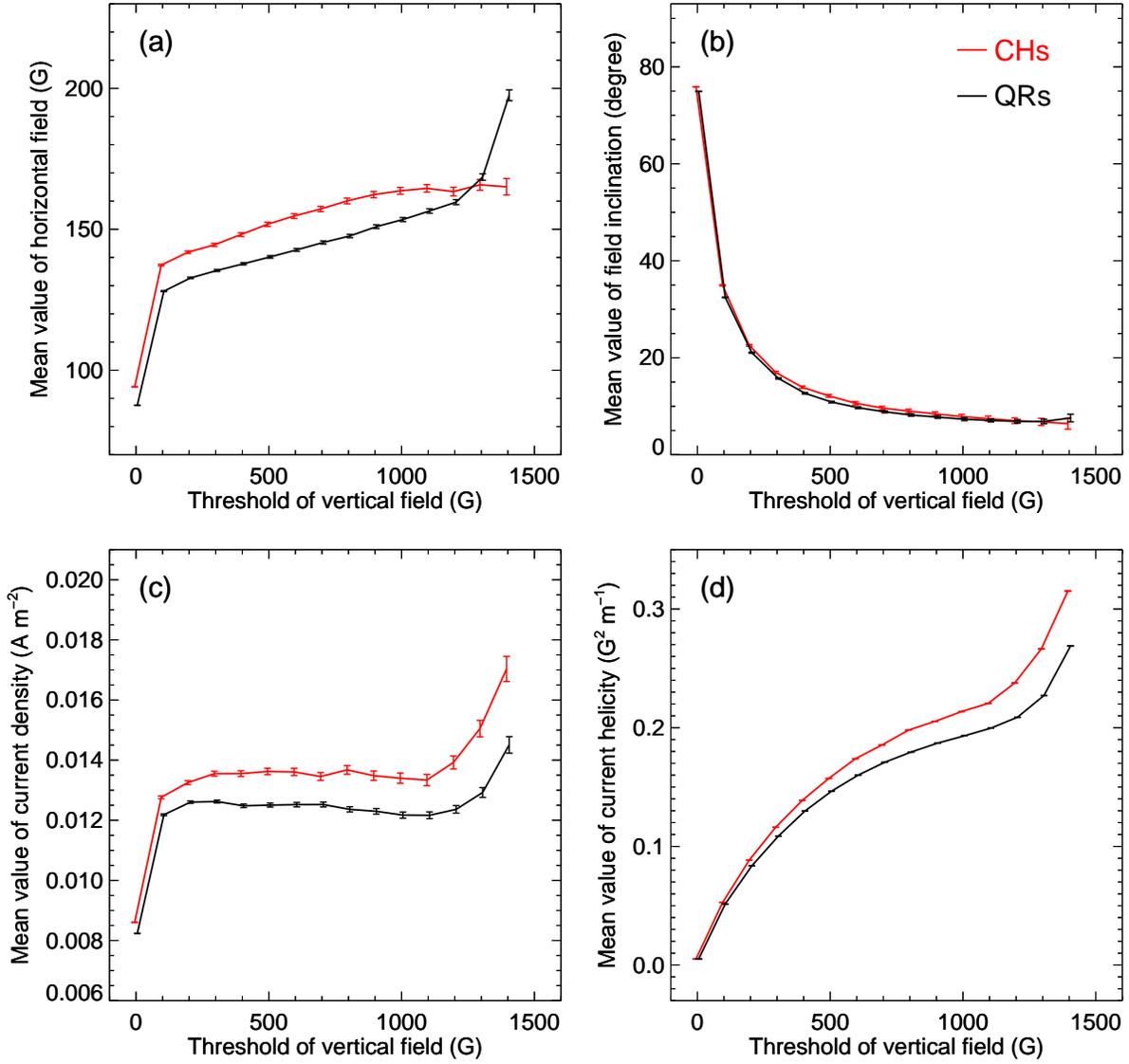} \caption{Similar to Figure 5 but
for the CHs (CH1 and CH2) and QRs (QR1 and QR2). \label{fig}}
\end{figure}
\clearpage

\end{document}